\documentclass[a4paper,12pt]{revtex4}

\usepackage{amsmath}
\usepackage{graphicx}
\usepackage{xcolor}
\usepackage[colorlinks=true, linkcolor=blue, citecolor=red]{hyperref}
\usepackage{enumitem} 
\usepackage[normalem]{ulem}

\providecommand{\abs}[1]{\left\lvert#1\right\rvert}

\newcommand{\ggF}{gg{\rm F}}
\newcommand{\VBF}{V{\rm BF}}
\newcommand{\GeV}{{\, \rm GeV}}

\begin{document}

\title{Is there new particle \\running in the loop-induced $H\gamma\gamma$ and $Hgg$ vertex?}

\author{Seungwon Baek}
\email{swbaek@kias.re.kr}
\affiliation{School of Physics, KIAS, 85 Hoegiro, Seoul 02455, Korea}

\author{Xing-Bo Yuan}
\email{xbyuan@yonsei.ac.kr}
\affiliation{Quantum Universe Center, KIAS, 85 Hoegiro, Seoul 02455, Korea\\Physics Division, National Center for Theoretical Sciences, Hsinchu 300, Taiwan}

\begin{abstract}
Loop-induced $Hgg$ and $H\gamma\gamma$ coupling play an important role in the Higgs production and decay at the LHC. These vertices can be affected by various New Physics contributions, including new particles, anomalous Yukawa couplings, and so on. We point out some ratios of the Higgs signal strengths could help disentangle the contributions of new particles from other sources of New Physics.
\end{abstract}

\maketitle

\newpage

\section{Introduction}\label{sec:introduction}
The first run of the LHC was very successful with the discovery of the Higgs boson~\cite{Aad:2012tfa, Chatrchyan:2012xdj}. The measured couplings of the observed Higgs boson to fermions and gauge bosons are compatible with the Standard Model (SM) predictions at a typical $\mathcal O(10-20)\%$ precision level~\cite{Khachatryan:2016vau}. The next step would be precision Higgs coupling measurements, which will be performed at the Run II of the LHC and its high-luminosity upgrade~\cite{ATLAS:2013hta,CMS:2013xfa}, as well as planned high-energy colliders~\cite{Baer:2013cma,CEPC-SPPCStudyGroup:2015csa}. These precision measurements will allow more complete understanding of Electro-Weak Symmetry Breaking (EWSB) and are crucial for searching for physics Beyond the Standard Model (BSM)~\cite{Csaki:2015hcd,Mariotti:2016owy}.

BSM physics can contribute to all the Higgs couplings, including the couplings to fermions $Hff$ and massive gauge
bosons $HVV$, which control various Higgs production and decay channels at the LHC. Unlike the tree-level couplings, the
Higgs couplings to photon $H\gamma\gamma$ and gluon $Hgg$ are generated by loop diagrams in the SM. Therefore, they can
be affected by BSM physics in a variety of ways~\cite{Gunion:1989we,Dawson:2013bba}. Generally, we can divide the various BSM sources into two scenarios. In
the first scenario, the $Hgg$ and $H\gamma\gamma$ couplings are affected only through the loop diagrams in the SM, where
the $Hff$ or $HWW$ couplings are modified. In the second scenario, new particles also enter in the loop diagrams and
affect the $Hgg$ and $H\gamma\gamma$ couplings directly. Even if deviation of the rate for the gluon-gluon fusion production
or $h\to \gamma\gamma$ decay were observed in the future, it is not straightforward to determine whether the anomaly is
due to contributions from modified tree-level couplings or from new particles. 
One approach to probe these two different BSM sources would be to check if the SM
tree-level coupling scale factors are sufficient to fit the Higgs data~\cite{Heinemeyer:2013tqa}. Alternatively, in this paper, we will build ratios from the Higgs signal strengths to disentangle the contributions of new particles from other BSM sources. All possible
BSM contributions are parameterized in the framework of Effective Field Theory (EFT). Numerical analysis with the LHC
Run I data is also performed.

This paper is organized as follows: In Sec.~\ref{sec:framework}, BSM contributions to the Higgs production and decay
channels at the LHC are investigated in the EFT framework. Ratios of the Higgs signal strengths are built to analyze the
BSM contributions. We present our numerical results in Sec.~\ref{sec:analysis} and conclude in
Sec.~\ref{sec:conclusion}.

\section{Theoretical framework}\label{sec:framework}
Considering current Higgs measurements at the LHC, a general theoretical framework can be provided by the effective Lagrangian \cite{Carmi:2012in,Buchalla:2015wfa},
\begin{align}\label{eq:effective Lagrangian}
\mathcal L=&2c_V\bigl ( m_W^2 W^+_\mu W^{-\mu}+\frac{1}{2}m_Z^2Z_\mu Z^\mu \bigr )\frac{h}{v}+c_\gamma \frac{\alpha_e}{\pi } F_{\mu\nu}F^{\mu\nu} \frac{h}{v}+c_g \frac{\alpha_s}{12\pi} G_{\mu\nu}^a G^{a\mu\nu} \frac{h}{v}\nonumber\\
&-c_t \frac{m_t}{ v}\bar t t h -c_b \frac{m_b}{ v}\bar b b h -c_c \frac{m_c}{v} \bar c c h -c_\tau\frac{m_\tau}{ v}\bar\tau \tau h,
\end{align}
with $v=246\GeV$. In the SM, the coupling $c_f$ ($f=c,b,t,\tau$) and $c_V$ ($V=W,Z$) are generated by tree-level diagrams and
equal to unity, while $c_\gamma$ and $c_g$ arise at loop level and vanish here by definition. Note that 
$c_\gamma$ and $c_g$ are reserved only for new particle contributions. The contributions from the $Hff$ or $HVV$ are added separately. Generally, all these
couplings could deviate from their SM values due to BSM physics. It is convenient to define the following effective
couplings
  \begin{align}\label{eq:effective coupling}
    C_f &= c_f, \nonumber\\
    C_V &= c_V, \nonumber\\
    C_\gamma & = c_\gamma - \frac{1}{8} \biggl ( \sum_f c_f N_{c,f} Q_f^2 \mathcal A_{\frac{1}{2}}(x_f) +c_V \mathcal A_1 (x_W) \biggr ), \nonumber\\
    C_g & = c_g -\frac{3}{4} \sum_q c_q \mathcal A_{\frac{1}{2}}(x_q),
  \end{align}
where $x_f = 4m_f^2/m_h^2$, and the normalized ones
\begin{align}\label{eq:normalized coupling}
\kappa_i = C_i/ C_i^{\rm SM}.
\end{align}
The $C_i^{\rm SM}$'s can be obtained from (\ref{eq:effective coupling}) by setting $c_f=c_V=1, c_\gamma=c_g=0$, and it
turns out that $C_f^{\rm SM}=C_V^{\rm SM}=1$,  $C_\gamma^{\rm SM} =-0.81$, and $C_g^{\rm SM} =0.99 +0.07 i$.
Here the one-loop functions read~\cite{Gunion:1989we,Djouadi:2005gi}
\begin{align}
  \mathcal A_{\frac{1}{2}} (x)&= -2x\left[ 1+(1-x)f(x) \right], \nonumber\\
  \mathcal A_{1} (x)  &= 2 + 3x + 3x(2-x)f(x),
\end{align}
where
\begin{align}
  f(x)=
  \begin{cases}
   \left[\sin^{-1}(1/\sqrt{x}) \right]^2, & x \ge 1,\\
    -\frac{1}{4}\left[\log\left(\frac{1+\sqrt{1-x}}{1-\sqrt{1-x}}\right)-i\pi\right]^2, & x<1.
  \end{cases}
\end{align}
These effective couplings describe the interactions of the Higgs boson to the SM fermions and gauge bosons, and can be straightforwardly used to investigate the Higgs observables, such as $\Gamma(h \to \gamma \gamma)  \propto \left \lvert C_\gamma \right\rvert^2 $ and $\sigma_{\ggF}  \propto \left \lvert C_g \right \rvert^2$. Generally, numerical results of these effective couplings are 
\begin{align}
C_\gamma &= c_\gamma + 0.23 c_t - 1.04 c_V, \nonumber\\
C_g &= c_g + 1.04c_t - (0.05 - 0.07i) c_b,
\end{align}
where the contributions from the light fermions $c$ and $\tau$ are less than 1\% and have been neglected. Fitting of $C_i$ to various experimental data has been performed by several groups~\cite{Belanger:2012gc,Belanger:2013xza,Ellis:2013lra,Giardino:2013bma,Fichet:2015xla} to test the SM and also to search for the BSM.

In the BSM, the effective couplings $C_i$ in eq.~\eqref{eq:effective coupling} may deviate from the SM values. Generally there are two scenarios:
\begin{description}[leftmargin = 2em, labelindent = 0em] 
\item[Scenario I:] $c_f,c_V\neq 1$ and $c_\gamma,c_g=0$ \newline Among the couplings $c_i$ in eq.~\eqref{eq:effective Lagrangian}, only the Higgs couplings to the fermions and massive gauge bosons $c_f$ and $c_V$ are directly modified by the BSM contributions. Therefore, the BSM effects on the effective couplings to gluon and photon are implemented through the loop diagrams in the SM with the modified $c_f$ and $c_V$, whose numerical results read
\begin{align}
C_\gamma &= 0.23 c_t - 1.04 c_V, \nonumber\\
C_g &=  1.04c_t - (0.05 - 0.07i) c_b.
\end{align}
This scenario usually corresponds to the BSM models with large extra dimension~\cite{Maru:2016vwr} or BSM candidates without
new electromagnetically charged particles coupled to Higgs such as gauged $L_\mu-L_\tau$ extensions of the SM~\cite{Baek:2001kca,Baek:2015fea}.

\item[Scenario II:] $c_f,c_V\neq 1$ and $c_\gamma,c_g \neq 0$\\
In some BSM candidates, new electrically charged or colored particles couple to the Higgs boson and generate loop-induced $H\gamma\gamma$ or $Hgg$ couplings, which give rise to nonzero $c_\gamma$ or $c_g$. In this scenario, the effective Higgs couplings to gluon and photon can be generally written as
\begin{align}
C_\gamma &= c_\gamma + 0.23 c_t - 1.04 c_V, \nonumber\\
C_g &= c_g + 1.04c_t - (0.05 - 0.07i) c_b.
\end{align}
This scenario is the most general case, since all the couplings $c_i$ could deviate from their SM
values~\cite{Almeida:2012bq,Baek:2015mea,Baek:2015fma}. 
It is noted that nonzero $c_\gamma$ can also be generated by the anomalous triple gauge coupling $W^+W^-\gamma$, which, however, is strongly constrained at high-energy colliders~\cite{Schael:2013ita,Aaltonen:2015hza,Aaltonen:2016pos,Aad:2016wpd,Khachatryan:2015sga,Falkowski:2015jaa,Falkowski:2016cxu} or from the $B$ and $K$ meson decays~\cite{He:1993hx,Bobeth:2015zqa}.
\end{description}
In the following, we will build some observables to discriminate these two scenarios.

At the LHC, Higgs boson production mainly occurs through the following channels,
\begin{center}
\begin{tabular}{r l r l}
gluon-gluon fusion ($\ggF$)  &: $gg\to h$ & \qquad\qquad coupling  &: $C_g$\\
vector boson fusion ($\VBF$)  &: $qq\to qqh$ &  coupling  &: $C_V$\\
associated production ($VH$)  &: $q q \to W/Z h$ &  coupling  &: $C_V$  \\
associated production ($ttH$)  &: $qq/gg \to tth$ & coupling  &: $C_t$,
\end{tabular}
\end{center}
where the effective couplings relevant for each process are also listed. It is noted that, associated production with a $Z$ boson also includes a process $gg \to Zh$ induced by top quark loops, which are affected by the effective coupling $C_t$~\cite{Khachatryan:2016vau}. For simplicity, we only consider the associated production with a $W$ boson in the following. In this work, the most relevant decay modes of the Higgs boson are listed below:
\begin{center}
\begin{tabular}{r l r l}
decay &: $h \to \gamma\gamma$ & \qquad \qquad coupling & : $C_\gamma$ \\
decay &: $h \to WW^*/ZZ^*$ & coupling & : $C_V$ \\
decay &: $h \to bb / \tau\tau$ & coupling & : $C_b,C_\tau$
\end{tabular}
\end{center}
The SM predictions on the Higgs production cross sections and decay branching ratios have been summarized in refs.~\cite{Dittmaier:2011ti,Dittmaier:2012vm,Heinemeyer:2013tqa}. 

To characterise the Higgs boson property, a signal strength is usually defined for a particular Higgs production and decay process $i \to h \to j$ at the LHC~\cite{Khachatryan:2016vau},
  \begin{align}
  \mu_i = \frac{\sigma_i}{\sigma_i^{\rm SM}}, \qquad   \mu^j = \frac{\mathcal B_j}{\mathcal B_j^{\rm SM}}, \qquad \mu_i^j = \mu_i \cdot \mu^j
  \end{align}
where $\sigma_i$ ($i = \ggF, \VBF, WH, ZH, ttH$) and $\mathcal B_j$ ($j=ZZ^*,WW^*,\gamma\gamma,\tau\tau,bb$) are the
production cross section for $i \to h$ and the branching ratio of the decay $h \to j$, respectively. Observables with
the superscript ``SM'' indicate their SM predictions. Therefore, $\mu_i=1$, $\mu^j=1$ and $\mu_i^j=1$ in the SM. It is
straightforward to represent the signal strengths with the normalized couplings defined in eq.~\eqref{eq:normalized
  coupling}, which read
\begin{align}\label{eq: mu: production}
\mu_{\ggF}=\abs{\kappa_g}^2, \qquad \mu_{\VBF} = \abs{\kappa_V}^2, \qquad \mu_{WH} = \abs{\kappa_V}^2, \qquad \mu_{ttH} = \abs{\kappa_t}^2,
\end{align}
and 
\begin{align}\label{eq: mu: decay}
\mu^{VV} = \frac{\abs{\kappa_V}^2}{r_\Gamma}, \qquad \mu^{\gamma\gamma} = \frac{\abs{\kappa_\gamma}^2}{r_\Gamma}, \qquad \mu^{ff}=\frac{\abs{\kappa_{f}}^2}{ r_\Gamma},
\end{align}
with $r_\Gamma=\Gamma_H/\Gamma_H^{\rm SM}$. Here, $\Gamma_H$ denotes the total decay width of the Higgs boson and
depends on all the couplings in the effective Lagrangian eq.~\eqref{eq:effective Lagrangian}.

Instead of the signal strengths themselves, it is interesting to consider their ratios, which can get rid of the total Higgs decay width~\cite{Zeppenfeld:2000td,Djouadi:2012rh,Djouadi:2015aba,Gunion:2012gc,Gunion:2012he,Ferreira:2012nv,Grossman:2013pt,Chang:2012ve,Goertz:2013kp}. Among all possible combinations, ratios of the signal strengths with same production or decay mode are the simplest, such as the ratios defined below
\begin{align}
 r_{1,V}^j = \frac{\mu_{\ggF}^j}{\mu_V^j},\qquad 
 r_{2,i}  = \frac{\mu_i^{\gamma\gamma}}{\mu_i^{VV}}, \qquad
 r_{3}^j  = \frac{\mu_{\ggF}^j}{\mu_{ttH}^j},
\end{align}
where $V$ denotes the $\VBF$ or $WH$ production mode. Since either the decay or production signal strength is canceled in each ratio, we can reduce them to the following basic ones
\begin{align}
  r_1 = \frac{\mu_{\ggF}}{\mu_V},\qquad
  r_2 = \frac{\mu^{\gamma\gamma}}{\mu^{VV}},\qquad
  r_3 = \frac{\mu_{\ggF}}{\mu_{ttH}}.
\end{align}
With eq.~\eqref{eq: mu: production} and \eqref{eq: mu: decay}, it's easy to obtain
\begin{align}
  r_1 = r_{1,V}^j=\abs{\frac{\kappa_g}{\kappa_V}}^2, \qquad
  r_2 = r_{2,i}=\abs{\frac{\kappa_\gamma}{\kappa_V}}^2,\qquad
  r_3 = r_3^j =\abs{\frac{\kappa_g}{\kappa_t}}^2.
\end{align}
In \textit{scenarios I} and \textit{II}, their numerical results read
\begin{align}\label{eq:r:numerical results}
  &        && \quad\text{Scenario I} && \quad\text{Scenario II} \nonumber\\
  &r_1= && \abs{1.06\frac{c_t}{c_V}}^2, && \abs{1.06\frac{c_t}{c_V}+1.03\frac{c_g}{c_V}}^2,\nonumber\\
  &r_2= && \abs{-1.27+0.28\frac{c_t}{c_V}}^2, && \abs{-1.27 + 0.28\frac{c_t}{c_V} +1.2 \frac{c_\gamma}{c_V}}^2, \nonumber\\
  &r_3= && \abs{1.06}^2, && \abs{1.06 + 1.03 \frac{c_g}{c_t}}^2,
\end{align}
where the contribution from $b$ quark has been neglected. In \textit{scenario I}, the ratios $r_1$ and $r_2$ only depend
on one parameter $c_t/c_V$, and the ratio $r_3$ is a constant. So these three ratios are strongly correlated with each
other. In \textit{scenario II}, however, the ratios receive additional contributions from $c_\gamma$ or $c_g$, and the
correlation between them no longer exists. Therefore, it is possible to distinguish between \textit{scenario I} and
\textit{II} by investigating the correlations between the ratios $r_1$, $r_2$ and $r_3$.

Before presenting the numerical analysis, a few comments are given here:
\begin{itemize}
\item When obtaining the numerical results of $r_1$, $r_2$ and $r_3$ in eq.~\eqref{eq:r:numerical results}, the
  contribution from $b$ quark is neglected. Since it only accounts for around 10\% of the effective coupling of the
  Higgs boson to gluon $C_g$, the one-parameter correlation between $r_1$, $r_2$ and $r_3$ in \textit{scenario I} is not
  polluted so much even after including this contribution. Furthermore, in the case that the Yukawa couplings $c_f$ are
  universal, the correlations in \textit{scenario I} hold exactly.
\item As advocated in refs.~\cite{Zeppenfeld:2000td,Djouadi:2012rh}, theoretical uncertainties in the ratios $r_i$ are largely eliminated, since either same production or decay channels are taken. Furthermore, some systematic errors, such as the one related with luminosity measurements, also cancel out in the ratios.
\item In the effective Lagrangian eq.~\eqref{eq:effective Lagrangian}, the BSM contributions to the tree-level $Hff$ and $HVV$ vertices are equivalent to multiplicative overall factors. Therefore, their QCD corrections cancel in the relevant signal strengths. For loop-induced vertices, the QCD corrections have been computed up to $\text{N}^3\text{LO}$ for $gg \to h$~\cite{Chetyrkin:1997un,Kramer:1996iq,Schroder:2005hy,Chetyrkin:2005ia,Baikov:2006ch,Inami:1982xt,Djouadi:1991tka,Chetyrkin:1997iv,Anastasiou:2015ema,Spira:1995rr} and NLO for $h \to \gamma\gamma$~\cite{Spira:1995rr,Djouadi:1993ji,Zheng:1990qa,Djouadi:1990aj,Dawson:1992cy,Melnikov:1993tj,Inoue:1994jq,Fleischer:2004vb}, and the NLO EW corrections are also available~\cite{Aglietti:2004nj,Aglietti:2004ki,Actis:2008ts,Degrassi:2004mx,Actis:2008ug,Degrassi:2005mc}. Although they modify the effective couplings $C_g$ and $C_\gamma$ by around 10\% level, their effects appear in both the denominator and numerator and are expected to be largely canceled in the signal strengths $\mu_{\ggF}$ and $\mu^{\gamma\gamma}$. We leave the high-order QCD and EW corrections to the ratios $r_{1-3}$ for future work.
\item It is possible to define other ratios which present similar features as $r_{1-3}$, such as
  $\mu_{\ggF}^{VV}/\mu_{V}^{\gamma\gamma}$. For simplicity, these ratios are not included in this paper.
\end{itemize}

\section{Numerical analysis}\label{sec:analysis}
\begin{figure}[t]
  \centering
  \includegraphics[width=0.32\textwidth]{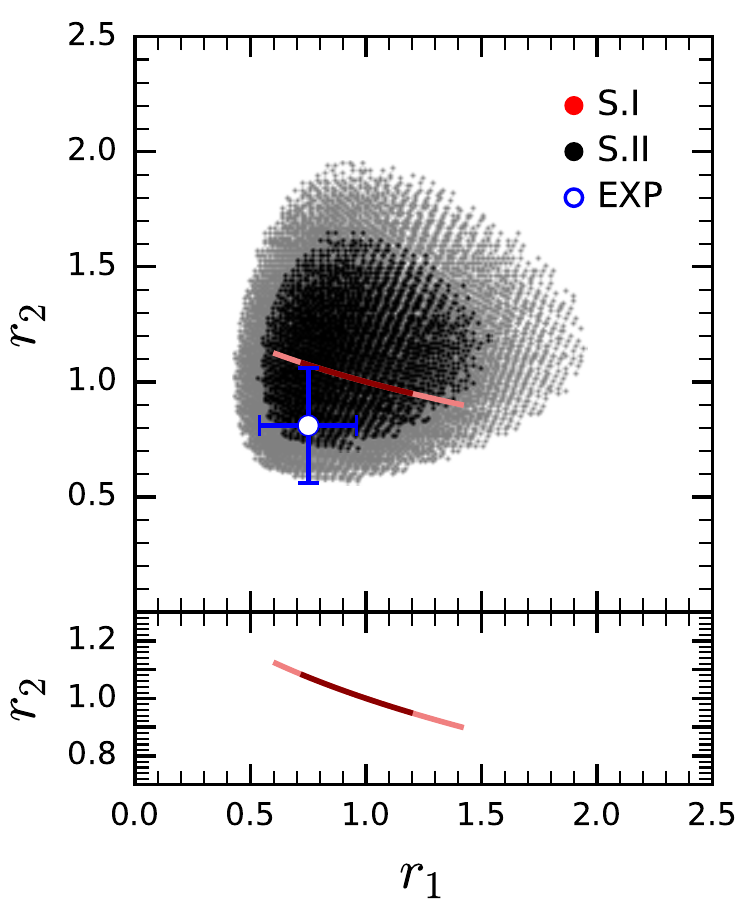}
  \includegraphics[width=0.32\textwidth]{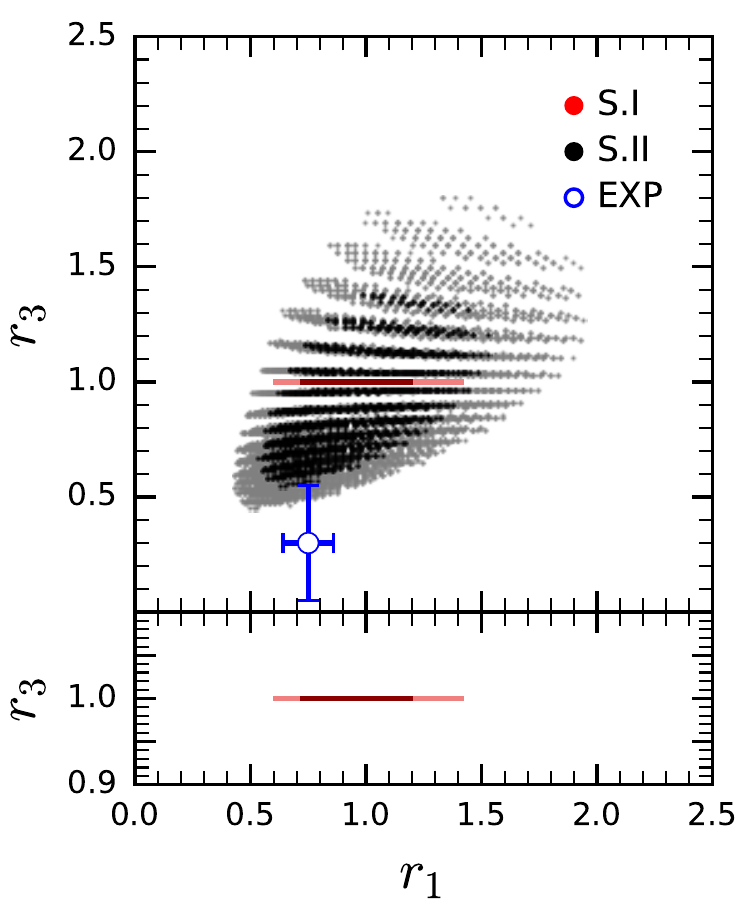}
  \includegraphics[width=0.32\textwidth]{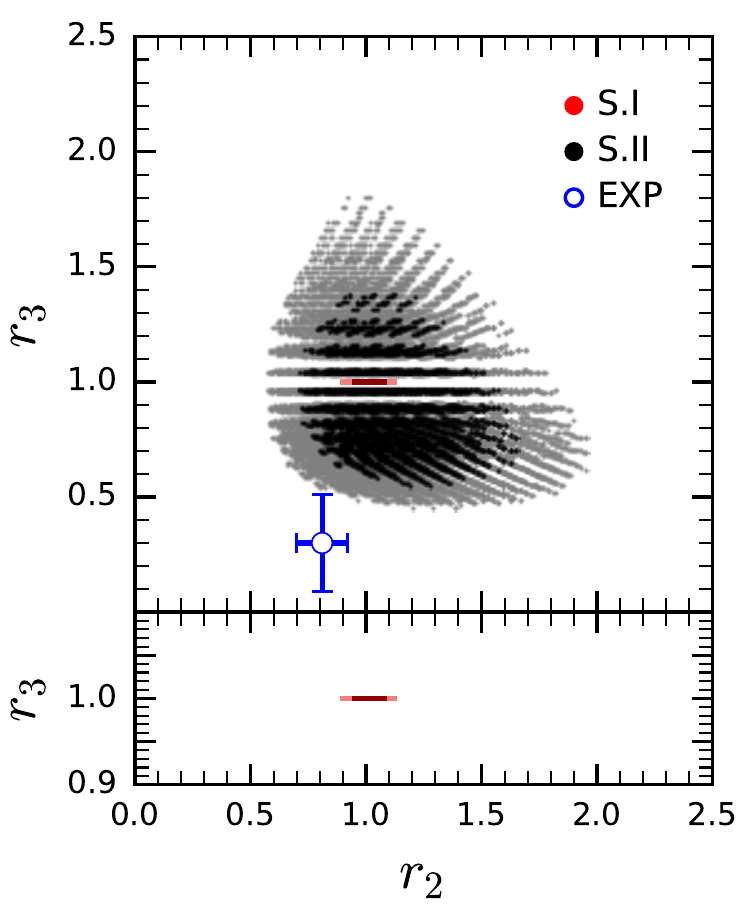}
  \caption{\baselineskip 3ex
    Correlations between the ratios $r_1$, $r_2$ and $r_3$ in the case of universal scale factors for the Yukawa couplings, which are obtained from the allowed parameter space S.I $(c_V, c_F)$ (dark red: 68\% CL, light red: 95\% CL) and S.II $(c_V, c_F, c_\gamma, c_g)$ (black: 68\% CL, gray: 95\% CL). The experimental data with $1\,\sigma$ error is illustrated by the blue point. Scaled plot for $r_2$ and $r_3$ are also shown below each figure.
}
\label{fig:correlation:universal}
\end{figure}

\begin{figure}[t]
  \centering
  \includegraphics[width=0.32\textwidth]{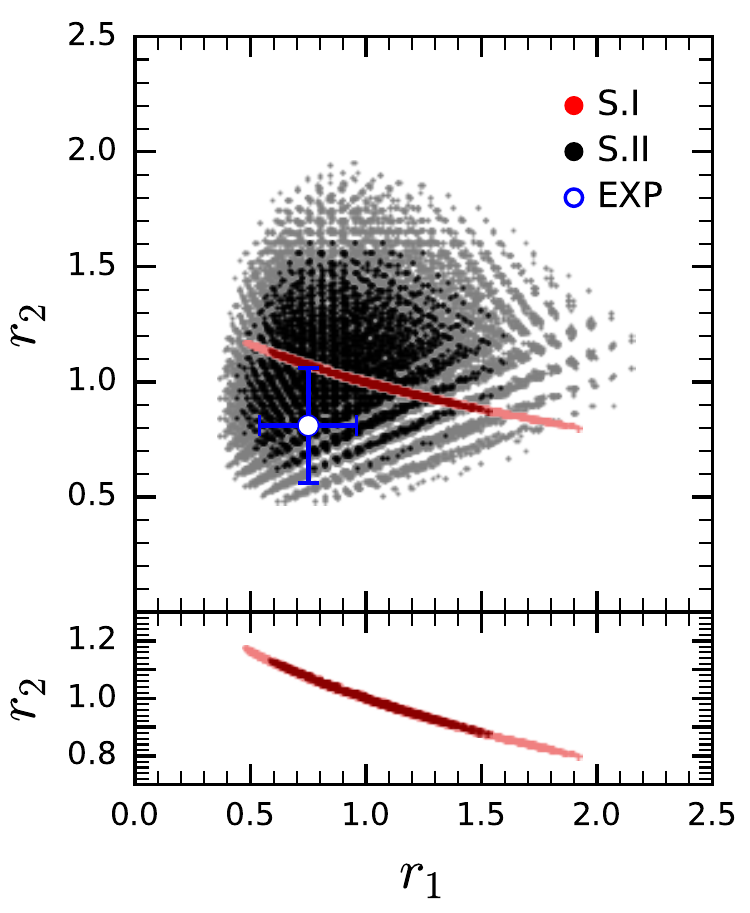}
  \includegraphics[width=0.32\textwidth]{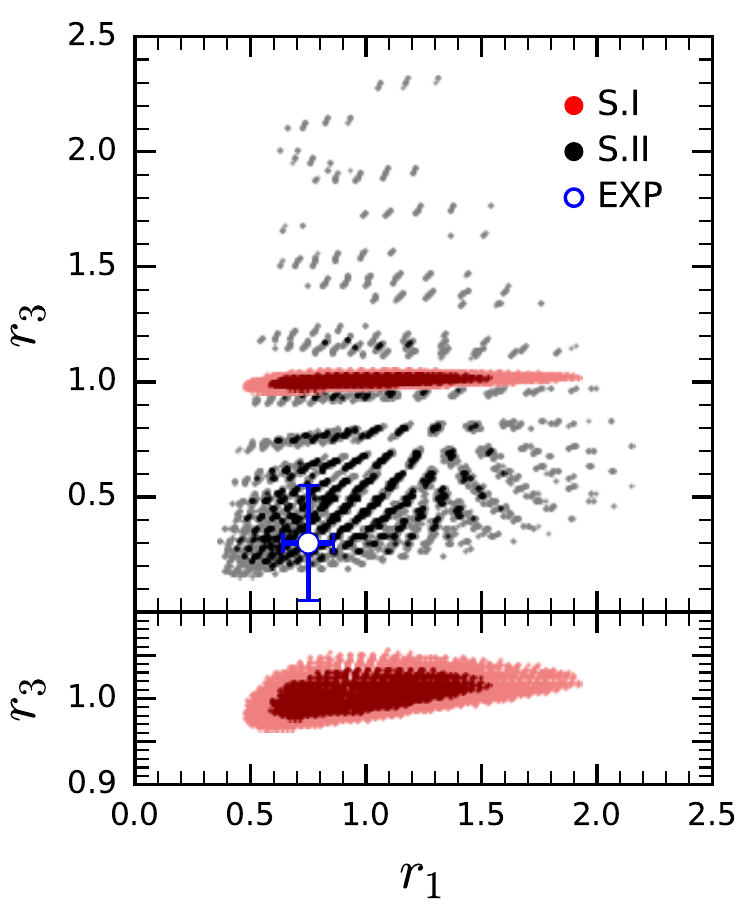}
  \includegraphics[width=0.32\textwidth]{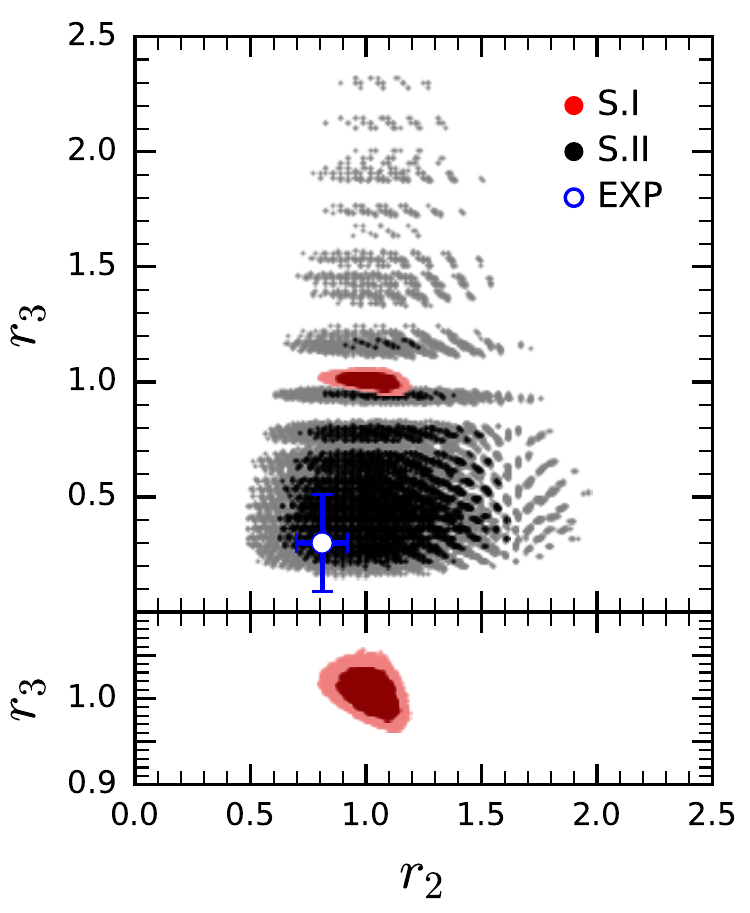}
  \caption{\baselineskip 3ex
    As in Fig.~\ref{fig:correlation:universal}, but from the parameter space S.I $(c_V, c_b, c_t, c_\tau)$ and S.II $(c_V, c_b, c_t, c_\tau, c_\gamma, c_g)$.
}
\label{fig:correlation:non-universal}
\end{figure}
As discussed in the previous section, the correlations between the ratios $r_i$ in \textit{scenario I} are slightly
polluted by the contributions from light fermions. To account for such contributions, we perform numerical analysis in
two cases: the scale factors for the Yukawa couplings are universal or non-universal, which correspond to the following
parameter sets:
\begin{center}
  \begin{tabular}{r l l }
       & universal coupling \qquad\qquad &      non-universal coupling  \\
     \textit{scenario \,\:I}\qquad\qquad & $(c_V, c_F)$ & $(c_V, c_b, c_t, c_\tau)$\\
     \textit{scenario II}\qquad\qquad& $(c_V, c_F, c_\gamma, c_g)$& $(c_V, c_b, c_t, c_\tau, c_\gamma, c_g)$.\\
  \end{tabular}
\end{center}
We perform a global fit for each parameter set. The package \texttt{Lilith-1.1.3} with database \texttt{DB
  15.09}~\cite{Bernon:2015hsa} is used to take into account the Higgs signal strengths measured by LHC Run
I~\cite{ATLAS:2015,CMS:2015kwa} and Tevatron~\cite{Aaltonen:2013ioz}. Our parameter scan doesn't include the LHC Run II
data, since the corresponding Higgs signal strengths haven't been fully available
yet~\cite{Gemme:2016wgs,PalenciaCortezon:2016yiz}. Our global fit shows that $\mathcal O(20\%)$ deviations from the SM
values are allowed in the fit of $(c_V,c_F)$, while $\mathcal O(50\%)$ in the other three fits. The fit results for the
parameter set $(c_V,c_F)$ and $(c_V,c_b,c_t,c_\tau,c_\gamma,c_g)$ can also been found in ref.~\cite{Khachatryan:2016vau}
and \cite{Buchalla:2015qju}, respectively.

We can obtain the experimental values of $r_1$, $r_2$ and $r_3$ from the following ratios.
\begin{center}
  \begin{tabular}{r c c c c c c}
    & \hspace{2em} & $\sigma_{\VBF}/\sigma_{\ggF}$ & \hspace{2em} & $\mathcal B^{\gamma\gamma}/\mathcal B^{ZZ}$ & \hspace{2em} & $\sigma_{ttH}/\sigma_{ggF}$ \\
SM\,~\cite{Heinemeyer:2013tqa} && $0.082\pm 0.009$ &  & $0.0854\pm 0.0010$ &  & $0.0067 \pm 0.0010$\\
EXP~\cite{Khachatryan:2016vau} && $0.109\pm 0.034$ &  & $0.069\;\: \pm 0.018 \;\:$ & & $0.022\;\:\pm0.007\;\:$\\
  \end{tabular}
\end{center}
To be conservative, experimental errors have been symmetrized. Compared to the SM predictions, the current experimental uncertainties are rather large, which are mainly due to the limited statistics available~\cite{Khachatryan:2016vau} and expected to be improved significantly at the future LHC.

In the case of universal scale factors for the Yukawa couplings, the ratios $r_i$ in \textit{scenario I} and \textit{II}
are shown in Fig.~\ref{fig:correlation:universal}. In \textit{scenario I}, as expected, $r_1$ and $r_2$ is strongly
correlated with each other and $r_3$ is kept a constant.  However, the three ratios can vary independently in a wide
range in \textit{scenario II}. The numerical results in the case of the non-universal scale factors are shown in
Fig.~\ref{fig:correlation:non-universal}. Although the contributions from light fermions are included generally, the
correlation between $r_1$ and $r_2$  in \textit{scenario I} is still very strong and $r_3$ can depart from its SM value
by at most $5\%$. Considering the LHC Run I data, the measured ratios prefer \textit{scenario II},
but with large uncertainty.
At the future LHC, any observed deviation from the correlations in \textit{scenario I} would indicate
non-vanishing coupling $c_\gamma$ or $c_g$, which can provide a hint of new particles running in the loop of the
$H\gamma\gamma$ or $Hgg$ coupling.

\section{Conclusions}\label{sec:conclusion}
In this paper, we have investigated how to use the Higgs observables at the LHC to determine if new particles enter 
the loops of the $H\gamma\gamma$ and $H gg$ effective couplings. Accordingly, we considered two general BSM scenarios. In {\it
  scenario I}, only tree-level Higgs couplings to fermions and massive gauge bosons are affected. They can still change
the $H\gamma\gamma$ and $Hgg$ couplings through the corresponding SM loop diagrams. In {\it scenario II}, besides the
modified tree-level couplings, there are new particles which can enter in the loop-induced vertices and they modify the
$H\gamma\gamma$ and $Hgg$ couplings directly.

Effects of these two BSM scenarios on the Higgs observables at the LHC have been investigated in the EFT framework. We
have constructed ratios of the Higgs signal strengths to distinguish between the two scenarios. We find that, in
\textit{scenario I}, the ratios $r_1 = \mu_{\ggF}/\mu_V$ and $r_2 = \mu^{\gamma\gamma}/\mu^{VV}$ strongly correlate with
each other and $r_3 = \mu_{\ggF}/\mu_{ttH}$ can deviate from its SM value by at most $5\%$. However, the ratios in
\textit{scenario II} vary independently in a wide range. Due to these different features, the ratios can
    provide information on whether there are new particles running in the loops of $Hgg$ and $H\gamma\gamma$ vertices. With
    large statistics expected at the future LHC, these ratios can be determined with a high precision, which makes them
    powerful tools to analysis possible anomalous Higgs couplings.





\section*{Acknowledgments}
SB is supported in part by National Research Foundation of Korea (NRF) Research Grant NRF-2015R1A2A1A05001869. XY is supported by NCTS. XY thanks Xiao-Gang He, Yun Jiang and Pyungwon Ko for useful discussions, and KIAS and QUC for its hospitality, where this work was mainly conducted. We thank J$\acute{\text{e}}$r$\acute{\text{e}}$my Bernon for explaining the \texttt{Lilith} code.

\bibliography{ref}

\begin{thebibliography}{76}
\expandafter\ifx\csname natexlab\endcsname\relax\def\natexlab#1{#1}\fi
\expandafter\ifx\csname bibnamefont\endcsname\relax
  \def\bibnamefont#1{#1}\fi
\expandafter\ifx\csname bibfnamefont\endcsname\relax
  \def\bibfnamefont#1{#1}\fi
\expandafter\ifx\csname citenamefont\endcsname\relax
  \def\citenamefont#1{#1}\fi
\expandafter\ifx\csname url\endcsname\relax
  \def\url#1{\texttt{#1}}\fi
\expandafter\ifx\csname urlprefix\endcsname\relax\def\urlprefix{URL }\fi
\providecommand{\bibinfo}[2]{#2}
\providecommand{\eprint}[2][]{\url{#2}}

\bibitem[{\citenamefont{Aad et~al.}(2012)}]{Aad:2012tfa}
\bibinfo{author}{\bibfnamefont{G.}~\bibnamefont{Aad}} \bibnamefont{et~al.}
  (\bibinfo{collaboration}{ATLAS}), \bibinfo{journal}{Phys. Lett.}
  \textbf{\bibinfo{volume}{B716}}, \bibinfo{pages}{1} (\bibinfo{year}{2012}),
  \eprint{1207.7214}.

\bibitem[{\citenamefont{Chatrchyan et~al.}(2012)}]{Chatrchyan:2012xdj}
\bibinfo{author}{\bibfnamefont{S.}~\bibnamefont{Chatrchyan}}
  \bibnamefont{et~al.} (\bibinfo{collaboration}{CMS}), \bibinfo{journal}{Phys.
  Lett.} \textbf{\bibinfo{volume}{B716}}, \bibinfo{pages}{30}
  (\bibinfo{year}{2012}), \eprint{1207.7235}.

\bibitem[{\citenamefont{Aad et~al.}(2016{\natexlab{a}})}]{Khachatryan:2016vau}
\bibinfo{author}{\bibfnamefont{G.}~\bibnamefont{Aad}} \bibnamefont{et~al.}
  (\bibinfo{collaboration}{ATLAS, CMS}), \bibinfo{journal}{JHEP}
  \textbf{\bibinfo{volume}{08}}, \bibinfo{pages}{045}
  (\bibinfo{year}{2016}{\natexlab{a}}), \eprint{1606.02266}.

\bibitem[{\citenamefont{Collaboration}(2013{\natexlab{a}})}]{ATLAS:2013hta}
\bibinfo{author}{\bibfnamefont{A.}~\bibnamefont{Collaboration}}
  (\bibinfo{collaboration}{ATLAS}), in \emph{\bibinfo{booktitle}{{Proceedings,
  Community Summer Study 2013: Snowmass on the Mississippi (CSS2013):
  Minneapolis, MN, USA, July 29-August 6, 2013}}}
  (\bibinfo{year}{2013}{\natexlab{a}}), \eprint{1307.7292},
  \urlprefix\url{http://inspirehep.net/record/1245017/files/arXiv:1307.7292.pdf}.

\bibitem[{\citenamefont{Collaboration}(2013{\natexlab{b}})}]{CMS:2013xfa}
\bibinfo{author}{\bibfnamefont{C.}~\bibnamefont{Collaboration}}
  (\bibinfo{collaboration}{CMS}), in \emph{\bibinfo{booktitle}{{Proceedings,
  Community Summer Study 2013: Snowmass on the Mississippi (CSS2013):
  Minneapolis, MN, USA, July 29-August 6, 2013}}}
  (\bibinfo{year}{2013}{\natexlab{b}}), \eprint{1307.7135},
  \urlprefix\url{http://inspirehep.net/record/1244669/files/arXiv:1307.7135.pdf}.

\bibitem[{\citenamefont{Baer et~al.}(2013)\citenamefont{Baer, Barklow, Fujii,
  Gao, Hoang, Kanemura, List, Logan, Nomerotski, Perelstein
  et~al.}}]{Baer:2013cma}
\bibinfo{author}{\bibfnamefont{H.}~\bibnamefont{Baer}},
  \bibinfo{author}{\bibfnamefont{T.}~\bibnamefont{Barklow}},
  \bibinfo{author}{\bibfnamefont{K.}~\bibnamefont{Fujii}},
  \bibinfo{author}{\bibfnamefont{Y.}~\bibnamefont{Gao}},
  \bibinfo{author}{\bibfnamefont{A.}~\bibnamefont{Hoang}},
  \bibinfo{author}{\bibfnamefont{S.}~\bibnamefont{Kanemura}},
  \bibinfo{author}{\bibfnamefont{J.}~\bibnamefont{List}},
  \bibinfo{author}{\bibfnamefont{H.~E.} \bibnamefont{Logan}},
  \bibinfo{author}{\bibfnamefont{A.}~\bibnamefont{Nomerotski}},
  \bibinfo{author}{\bibfnamefont{M.}~\bibnamefont{Perelstein}},
  \bibnamefont{et~al.} (\bibinfo{year}{2013}), \eprint{1306.6352}.

\bibitem[{\citenamefont{Group}(2015)}]{CEPC-SPPCStudyGroup:2015csa}
\bibinfo{author}{\bibfnamefont{C.-S.~S.} \bibnamefont{Group}}
  (\bibinfo{year}{2015}).

\bibitem[{\citenamefont{Csaki et~al.}(2016)\citenamefont{Csaki, Grojean, and
  Terning}}]{Csaki:2015hcd}
\bibinfo{author}{\bibfnamefont{C.}~\bibnamefont{Csaki}},
  \bibinfo{author}{\bibfnamefont{C.}~\bibnamefont{Grojean}}, \bibnamefont{and}
  \bibinfo{author}{\bibfnamefont{J.}~\bibnamefont{Terning}},
  \bibinfo{journal}{Rev. Mod. Phys.} \textbf{\bibinfo{volume}{88}},
  \bibinfo{pages}{045001} (\bibinfo{year}{2016}), \eprint{1512.00468}.

\bibitem[{\citenamefont{Mariotti and Passarino}(2016)}]{Mariotti:2016owy}
\bibinfo{author}{\bibfnamefont{C.}~\bibnamefont{Mariotti}} \bibnamefont{and}
  \bibinfo{author}{\bibfnamefont{G.}~\bibnamefont{Passarino}}
  (\bibinfo{year}{2016}), \eprint{1612.00269}.

\bibitem[{\citenamefont{Gunion et~al.}(2000)\citenamefont{Gunion, Haber, Kane,
  and Dawson}}]{Gunion:1989we}
\bibinfo{author}{\bibfnamefont{J.~F.} \bibnamefont{Gunion}},
  \bibinfo{author}{\bibfnamefont{H.~E.} \bibnamefont{Haber}},
  \bibinfo{author}{\bibfnamefont{G.~L.} \bibnamefont{Kane}}, \bibnamefont{and}
  \bibinfo{author}{\bibfnamefont{S.}~\bibnamefont{Dawson}},
  \bibinfo{journal}{Front. Phys.} \textbf{\bibinfo{volume}{80}},
  \bibinfo{pages}{1} (\bibinfo{year}{2000}).

\bibitem[{\citenamefont{Dawson et~al.}(2013)}]{Dawson:2013bba}
\bibinfo{author}{\bibfnamefont{S.}~\bibnamefont{Dawson}} \bibnamefont{et~al.},
  in \emph{\bibinfo{booktitle}{Proceedings, 2013 Community Summer Study on the
  Future of U.S. Particle Physics: Snowmass on the Mississippi (CSS2013):
  Minneapolis, MN, USA, July 29-August 6, 2013}} (\bibinfo{year}{2013}),
  \eprint{1310.8361}.

\bibitem[{\citenamefont{Andersen et~al.}(2013)}]{Heinemeyer:2013tqa}
\bibinfo{author}{\bibfnamefont{J.~R.} \bibnamefont{Andersen}}
  \bibnamefont{et~al.} (\bibinfo{collaboration}{LHC Higgs Cross Section Working
  Group}) (\bibinfo{year}{2013}), \eprint{1307.1347}.

\bibitem[{\citenamefont{Carmi et~al.}(2012)\citenamefont{Carmi, Falkowski,
  Kuflik, Volansky, and Zupan}}]{Carmi:2012in}
\bibinfo{author}{\bibfnamefont{D.}~\bibnamefont{Carmi}},
  \bibinfo{author}{\bibfnamefont{A.}~\bibnamefont{Falkowski}},
  \bibinfo{author}{\bibfnamefont{E.}~\bibnamefont{Kuflik}},
  \bibinfo{author}{\bibfnamefont{T.}~\bibnamefont{Volansky}}, \bibnamefont{and}
  \bibinfo{author}{\bibfnamefont{J.}~\bibnamefont{Zupan}},
  \bibinfo{journal}{JHEP} \textbf{\bibinfo{volume}{10}}, \bibinfo{pages}{196}
  (\bibinfo{year}{2012}), \eprint{1207.1718}.

\bibitem[{\citenamefont{Buchalla et~al.}(2015)\citenamefont{Buchalla, Cata,
  Celis, and Krause}}]{Buchalla:2015wfa}
\bibinfo{author}{\bibfnamefont{G.}~\bibnamefont{Buchalla}},
  \bibinfo{author}{\bibfnamefont{O.}~\bibnamefont{Cata}},
  \bibinfo{author}{\bibfnamefont{A.}~\bibnamefont{Celis}}, \bibnamefont{and}
  \bibinfo{author}{\bibfnamefont{C.}~\bibnamefont{Krause}},
  \bibinfo{journal}{Phys. Lett.} \textbf{\bibinfo{volume}{B750}},
  \bibinfo{pages}{298} (\bibinfo{year}{2015}), \eprint{1504.01707}.

\bibitem[{\citenamefont{Djouadi}(2008)}]{Djouadi:2005gi}
\bibinfo{author}{\bibfnamefont{A.}~\bibnamefont{Djouadi}},
  \bibinfo{journal}{Phys. Rept.} \textbf{\bibinfo{volume}{457}},
  \bibinfo{pages}{1} (\bibinfo{year}{2008}), \eprint{hep-ph/0503172}.

\bibitem[{\citenamefont{Belanger
  et~al.}(2013{\natexlab{a}})\citenamefont{Belanger, Dumont, Ellwanger, Gunion,
  and Kraml}}]{Belanger:2012gc}
\bibinfo{author}{\bibfnamefont{G.}~\bibnamefont{Belanger}},
  \bibinfo{author}{\bibfnamefont{B.}~\bibnamefont{Dumont}},
  \bibinfo{author}{\bibfnamefont{U.}~\bibnamefont{Ellwanger}},
  \bibinfo{author}{\bibfnamefont{J.~F.} \bibnamefont{Gunion}},
  \bibnamefont{and} \bibinfo{author}{\bibfnamefont{S.}~\bibnamefont{Kraml}},
  \bibinfo{journal}{JHEP} \textbf{\bibinfo{volume}{02}}, \bibinfo{pages}{053}
  (\bibinfo{year}{2013}{\natexlab{a}}), \eprint{1212.5244}.

\bibitem[{\citenamefont{Belanger
  et~al.}(2013{\natexlab{b}})\citenamefont{Belanger, Dumont, Ellwanger, Gunion,
  and Kraml}}]{Belanger:2013xza}
\bibinfo{author}{\bibfnamefont{G.}~\bibnamefont{Belanger}},
  \bibinfo{author}{\bibfnamefont{B.}~\bibnamefont{Dumont}},
  \bibinfo{author}{\bibfnamefont{U.}~\bibnamefont{Ellwanger}},
  \bibinfo{author}{\bibfnamefont{J.~F.} \bibnamefont{Gunion}},
  \bibnamefont{and} \bibinfo{author}{\bibfnamefont{S.}~\bibnamefont{Kraml}},
  \bibinfo{journal}{Phys. Rev.} \textbf{\bibinfo{volume}{D88}},
  \bibinfo{pages}{075008} (\bibinfo{year}{2013}{\natexlab{b}}),
  \eprint{1306.2941}.

\bibitem[{\citenamefont{Ellis and You}(2013)}]{Ellis:2013lra}
\bibinfo{author}{\bibfnamefont{J.}~\bibnamefont{Ellis}} \bibnamefont{and}
  \bibinfo{author}{\bibfnamefont{T.}~\bibnamefont{You}},
  \bibinfo{journal}{JHEP} \textbf{\bibinfo{volume}{06}}, \bibinfo{pages}{103}
  (\bibinfo{year}{2013}), \eprint{1303.3879}.

\bibitem[{\citenamefont{Giardino et~al.}(2014)\citenamefont{Giardino, Kannike,
  Masina, Raidal, and Strumia}}]{Giardino:2013bma}
\bibinfo{author}{\bibfnamefont{P.~P.} \bibnamefont{Giardino}},
  \bibinfo{author}{\bibfnamefont{K.}~\bibnamefont{Kannike}},
  \bibinfo{author}{\bibfnamefont{I.}~\bibnamefont{Masina}},
  \bibinfo{author}{\bibfnamefont{M.}~\bibnamefont{Raidal}}, \bibnamefont{and}
  \bibinfo{author}{\bibfnamefont{A.}~\bibnamefont{Strumia}},
  \bibinfo{journal}{JHEP} \textbf{\bibinfo{volume}{05}}, \bibinfo{pages}{046}
  (\bibinfo{year}{2014}), \eprint{1303.3570}.

\bibitem[{\citenamefont{Fichet and Moreau}(2016)}]{Fichet:2015xla}
\bibinfo{author}{\bibfnamefont{S.}~\bibnamefont{Fichet}} \bibnamefont{and}
  \bibinfo{author}{\bibfnamefont{G.}~\bibnamefont{Moreau}},
  \bibinfo{journal}{Nucl. Phys.} \textbf{\bibinfo{volume}{B905}},
  \bibinfo{pages}{391} (\bibinfo{year}{2016}), \eprint{1509.00472}.

\bibitem[{\citenamefont{Maru and Okada}(2016)}]{Maru:2016vwr}
\bibinfo{author}{\bibfnamefont{N.}~\bibnamefont{Maru}} \bibnamefont{and}
  \bibinfo{author}{\bibfnamefont{N.}~\bibnamefont{Okada}}
  (\bibinfo{year}{2016}), \eprint{1604.01150}.

\bibitem[{\citenamefont{Baek et~al.}(2001)\citenamefont{Baek, Deshpande, He,
  and Ko}}]{Baek:2001kca}
\bibinfo{author}{\bibfnamefont{S.}~\bibnamefont{Baek}},
  \bibinfo{author}{\bibfnamefont{N.~G.} \bibnamefont{Deshpande}},
  \bibinfo{author}{\bibfnamefont{X.~G.} \bibnamefont{He}}, \bibnamefont{and}
  \bibinfo{author}{\bibfnamefont{P.}~\bibnamefont{Ko}}, \bibinfo{journal}{Phys.
  Rev.} \textbf{\bibinfo{volume}{D64}}, \bibinfo{pages}{055006}
  (\bibinfo{year}{2001}), \eprint{hep-ph/0104141}.

\bibitem[{\citenamefont{Baek}(2016)}]{Baek:2015fea}
\bibinfo{author}{\bibfnamefont{S.}~\bibnamefont{Baek}}, \bibinfo{journal}{Phys.
  Lett.} \textbf{\bibinfo{volume}{B756}}, \bibinfo{pages}{1}
  (\bibinfo{year}{2016}), \eprint{1510.02168}.

\bibitem[{\citenamefont{Almeida et~al.}(2012)\citenamefont{Almeida, Bertuzzo,
  Machado, and Zukanovich~Funchal}}]{Almeida:2012bq}
\bibinfo{author}{\bibfnamefont{L.~G.} \bibnamefont{Almeida}},
  \bibinfo{author}{\bibfnamefont{E.}~\bibnamefont{Bertuzzo}},
  \bibinfo{author}{\bibfnamefont{P.~A.~N.} \bibnamefont{Machado}},
  \bibnamefont{and}
  \bibinfo{author}{\bibfnamefont{R.}~\bibnamefont{Zukanovich~Funchal}},
  \bibinfo{journal}{JHEP} \textbf{\bibinfo{volume}{11}}, \bibinfo{pages}{085}
  (\bibinfo{year}{2012}), \eprint{1207.5254}.

\bibitem[{\citenamefont{Baek and Nishiwaki}(2016)}]{Baek:2015mea}
\bibinfo{author}{\bibfnamefont{S.}~\bibnamefont{Baek}} \bibnamefont{and}
  \bibinfo{author}{\bibfnamefont{K.}~\bibnamefont{Nishiwaki}},
  \bibinfo{journal}{Phys. Rev.} \textbf{\bibinfo{volume}{D93}},
  \bibinfo{pages}{015002} (\bibinfo{year}{2016}), \eprint{1509.07410}.

\bibitem[{\citenamefont{Baek and Kang}(2016)}]{Baek:2015fma}
\bibinfo{author}{\bibfnamefont{S.}~\bibnamefont{Baek}} \bibnamefont{and}
  \bibinfo{author}{\bibfnamefont{Z.-F.} \bibnamefont{Kang}},
  \bibinfo{journal}{JHEP} \textbf{\bibinfo{volume}{03}}, \bibinfo{pages}{106}
  (\bibinfo{year}{2016}), \eprint{1510.00100}.

\bibitem[{\citenamefont{Schael et~al.}(2013)}]{Schael:2013ita}
\bibinfo{author}{\bibfnamefont{S.}~\bibnamefont{Schael}} \bibnamefont{et~al.}
  (\bibinfo{collaboration}{DELPHI, OPAL, LEP Electroweak, ALEPH, L3}),
  \bibinfo{journal}{Phys. Rept.} \textbf{\bibinfo{volume}{532}},
  \bibinfo{pages}{119} (\bibinfo{year}{2013}), \eprint{1302.3415}.

\bibitem[{\citenamefont{Aaltonen et~al.}(2015)}]{Aaltonen:2015hza}
\bibinfo{author}{\bibfnamefont{T.~A.} \bibnamefont{Aaltonen}}
  \bibnamefont{et~al.} (\bibinfo{collaboration}{CDF}), \bibinfo{journal}{Phys.
  Rev.} \textbf{\bibinfo{volume}{D91}}, \bibinfo{pages}{111101}
  (\bibinfo{year}{2015}), \bibinfo{note}{[Addendum: Phys.
  Rev.D92,no.3,039901(2015)]}, \eprint{1505.00801}.

\bibitem[{\citenamefont{Aaltonen et~al.}(2016)}]{Aaltonen:2016pos}
\bibinfo{author}{\bibfnamefont{T.~A.} \bibnamefont{Aaltonen}}
  \bibnamefont{et~al.} (\bibinfo{collaboration}{CDF}), \bibinfo{journal}{Phys.
  Rev.} \textbf{\bibinfo{volume}{D94}}, \bibinfo{pages}{032008}
  (\bibinfo{year}{2016}), \eprint{1606.06823}.

\bibitem[{\citenamefont{Aad et~al.}(2016{\natexlab{b}})}]{Aad:2016wpd}
\bibinfo{author}{\bibfnamefont{G.}~\bibnamefont{Aad}} \bibnamefont{et~al.}
  (\bibinfo{collaboration}{ATLAS}), \bibinfo{journal}{JHEP}
  \textbf{\bibinfo{volume}{09}}, \bibinfo{pages}{029}
  (\bibinfo{year}{2016}{\natexlab{b}}), \eprint{1603.01702}.

\bibitem[{\citenamefont{Khachatryan et~al.}(2016)}]{Khachatryan:2015sga}
\bibinfo{author}{\bibfnamefont{V.}~\bibnamefont{Khachatryan}}
  \bibnamefont{et~al.} (\bibinfo{collaboration}{CMS}), \bibinfo{journal}{Eur.
  Phys. J.} \textbf{\bibinfo{volume}{C76}}, \bibinfo{pages}{401}
  (\bibinfo{year}{2016}), \eprint{1507.03268}.

\bibitem[{\citenamefont{Falkowski
  et~al.}(2016{\natexlab{a}})\citenamefont{Falkowski, Gonzalez-Alonso, Greljo,
  and Marzocca}}]{Falkowski:2015jaa}
\bibinfo{author}{\bibfnamefont{A.}~\bibnamefont{Falkowski}},
  \bibinfo{author}{\bibfnamefont{M.}~\bibnamefont{Gonzalez-Alonso}},
  \bibinfo{author}{\bibfnamefont{A.}~\bibnamefont{Greljo}}, \bibnamefont{and}
  \bibinfo{author}{\bibfnamefont{D.}~\bibnamefont{Marzocca}},
  \bibinfo{journal}{Phys. Rev. Lett.} \textbf{\bibinfo{volume}{116}},
  \bibinfo{pages}{011801} (\bibinfo{year}{2016}{\natexlab{a}}),
  \eprint{1508.00581}.

\bibitem[{\citenamefont{Falkowski
  et~al.}(2016{\natexlab{b}})\citenamefont{Falkowski, Gonzalez-Alonso, Greljo,
  Marzocca, and Son}}]{Falkowski:2016cxu}
\bibinfo{author}{\bibfnamefont{A.}~\bibnamefont{Falkowski}},
  \bibinfo{author}{\bibfnamefont{M.}~\bibnamefont{Gonzalez-Alonso}},
  \bibinfo{author}{\bibfnamefont{A.}~\bibnamefont{Greljo}},
  \bibinfo{author}{\bibfnamefont{D.}~\bibnamefont{Marzocca}}, \bibnamefont{and}
  \bibinfo{author}{\bibfnamefont{M.}~\bibnamefont{Son}}
  (\bibinfo{year}{2016}{\natexlab{b}}), \eprint{1609.06312}.

\bibitem[{\citenamefont{He and McKellar}(1994)}]{He:1993hx}
\bibinfo{author}{\bibfnamefont{X.-G.} \bibnamefont{He}} \bibnamefont{and}
  \bibinfo{author}{\bibfnamefont{B.}~\bibnamefont{McKellar}},
  \bibinfo{journal}{Phys. Lett.} \textbf{\bibinfo{volume}{B320}},
  \bibinfo{pages}{165} (\bibinfo{year}{1994}), \eprint{hep-ph/9309228}.

\bibitem[{\citenamefont{Bobeth and Haisch}(2015)}]{Bobeth:2015zqa}
\bibinfo{author}{\bibfnamefont{C.}~\bibnamefont{Bobeth}} \bibnamefont{and}
  \bibinfo{author}{\bibfnamefont{U.}~\bibnamefont{Haisch}},
  \bibinfo{journal}{JHEP} \textbf{\bibinfo{volume}{09}}, \bibinfo{pages}{018}
  (\bibinfo{year}{2015}), \eprint{1503.04829}.

\bibitem[{\citenamefont{Dittmaier et~al.}(2011)}]{Dittmaier:2011ti}
\bibinfo{author}{\bibfnamefont{S.}~\bibnamefont{Dittmaier}}
  \bibnamefont{et~al.} (\bibinfo{collaboration}{LHC Higgs Cross Section Working
  Group}) (\bibinfo{year}{2011}), \eprint{1101.0593}.

\bibitem[{\citenamefont{Dittmaier et~al.}(2012)}]{Dittmaier:2012vm}
\bibinfo{author}{\bibfnamefont{S.}~\bibnamefont{Dittmaier}}
  \bibnamefont{et~al.} (\bibinfo{year}{2012}), \eprint{1201.3084}.

\bibitem[{\citenamefont{Zeppenfeld et~al.}(2000)\citenamefont{Zeppenfeld,
  Kinnunen, Nikitenko, and Richter-Was}}]{Zeppenfeld:2000td}
\bibinfo{author}{\bibfnamefont{D.}~\bibnamefont{Zeppenfeld}},
  \bibinfo{author}{\bibfnamefont{R.}~\bibnamefont{Kinnunen}},
  \bibinfo{author}{\bibfnamefont{A.}~\bibnamefont{Nikitenko}},
  \bibnamefont{and}
  \bibinfo{author}{\bibfnamefont{E.}~\bibnamefont{Richter-Was}},
  \bibinfo{journal}{Phys. Rev.} \textbf{\bibinfo{volume}{D62}},
  \bibinfo{pages}{013009} (\bibinfo{year}{2000}), \eprint{hep-ph/0002036}.

\bibitem[{\citenamefont{Djouadi}(2013)}]{Djouadi:2012rh}
\bibinfo{author}{\bibfnamefont{A.}~\bibnamefont{Djouadi}},
  \bibinfo{journal}{Eur. Phys. J.} \textbf{\bibinfo{volume}{C73}},
  \bibinfo{pages}{2498} (\bibinfo{year}{2013}), \eprint{1208.3436}.

\bibitem[{\citenamefont{Djouadi et~al.}(2016)\citenamefont{Djouadi, Quevillon,
  and Vega-Morales}}]{Djouadi:2015aba}
\bibinfo{author}{\bibfnamefont{A.}~\bibnamefont{Djouadi}},
  \bibinfo{author}{\bibfnamefont{J.}~\bibnamefont{Quevillon}},
  \bibnamefont{and}
  \bibinfo{author}{\bibfnamefont{R.}~\bibnamefont{Vega-Morales}},
  \bibinfo{journal}{Phys. Lett.} \textbf{\bibinfo{volume}{B757}},
  \bibinfo{pages}{412} (\bibinfo{year}{2016}), \eprint{1509.03913}.

\bibitem[{\citenamefont{Gunion et~al.}(2012)\citenamefont{Gunion, Jiang, and
  Kraml}}]{Gunion:2012gc}
\bibinfo{author}{\bibfnamefont{J.~F.} \bibnamefont{Gunion}},
  \bibinfo{author}{\bibfnamefont{Y.}~\bibnamefont{Jiang}}, \bibnamefont{and}
  \bibinfo{author}{\bibfnamefont{S.}~\bibnamefont{Kraml}},
  \bibinfo{journal}{Phys. Rev.} \textbf{\bibinfo{volume}{D86}},
  \bibinfo{pages}{071702} (\bibinfo{year}{2012}), \eprint{1207.1545}.

\bibitem[{\citenamefont{Gunion et~al.}(2013)\citenamefont{Gunion, Jiang, and
  Kraml}}]{Gunion:2012he}
\bibinfo{author}{\bibfnamefont{J.~F.} \bibnamefont{Gunion}},
  \bibinfo{author}{\bibfnamefont{Y.}~\bibnamefont{Jiang}}, \bibnamefont{and}
  \bibinfo{author}{\bibfnamefont{S.}~\bibnamefont{Kraml}},
  \bibinfo{journal}{Phys. Rev. Lett.} \textbf{\bibinfo{volume}{110}},
  \bibinfo{pages}{051801} (\bibinfo{year}{2013}), \eprint{1208.1817}.

\bibitem[{\citenamefont{Ferreira et~al.}(2013)\citenamefont{Ferreira, Santos,
  Haber, and Silva}}]{Ferreira:2012nv}
\bibinfo{author}{\bibfnamefont{P.~M.} \bibnamefont{Ferreira}},
  \bibinfo{author}{\bibfnamefont{R.}~\bibnamefont{Santos}},
  \bibinfo{author}{\bibfnamefont{H.~E.} \bibnamefont{Haber}}, \bibnamefont{and}
  \bibinfo{author}{\bibfnamefont{J.~P.} \bibnamefont{Silva}},
  \bibinfo{journal}{Phys. Rev.} \textbf{\bibinfo{volume}{D87}},
  \bibinfo{pages}{055009} (\bibinfo{year}{2013}), \eprint{1211.3131}.

\bibitem[{\citenamefont{Grossman et~al.}(2013)\citenamefont{Grossman, Surujon,
  and Zupan}}]{Grossman:2013pt}
\bibinfo{author}{\bibfnamefont{Y.}~\bibnamefont{Grossman}},
  \bibinfo{author}{\bibfnamefont{Z.}~\bibnamefont{Surujon}}, \bibnamefont{and}
  \bibinfo{author}{\bibfnamefont{J.}~\bibnamefont{Zupan}},
  \bibinfo{journal}{JHEP} \textbf{\bibinfo{volume}{03}}, \bibinfo{pages}{176}
  (\bibinfo{year}{2013}), \eprint{1301.0328}.

\bibitem[{\citenamefont{Chang et~al.}(2013)\citenamefont{Chang, Kang, Lee, Lee,
  Park, and Song}}]{Chang:2012ve}
\bibinfo{author}{\bibfnamefont{S.}~\bibnamefont{Chang}},
  \bibinfo{author}{\bibfnamefont{S.~K.} \bibnamefont{Kang}},
  \bibinfo{author}{\bibfnamefont{J.-P.} \bibnamefont{Lee}},
  \bibinfo{author}{\bibfnamefont{K.~Y.} \bibnamefont{Lee}},
  \bibinfo{author}{\bibfnamefont{S.~C.} \bibnamefont{Park}}, \bibnamefont{and}
  \bibinfo{author}{\bibfnamefont{J.}~\bibnamefont{Song}},
  \bibinfo{journal}{JHEP} \textbf{\bibinfo{volume}{05}}, \bibinfo{pages}{075}
  (\bibinfo{year}{2013}), \eprint{1210.3439}.

\bibitem[{\citenamefont{Goertz et~al.}(2013)\citenamefont{Goertz,
  Papaefstathiou, Yang, and Zurita}}]{Goertz:2013kp}
\bibinfo{author}{\bibfnamefont{F.}~\bibnamefont{Goertz}},
  \bibinfo{author}{\bibfnamefont{A.}~\bibnamefont{Papaefstathiou}},
  \bibinfo{author}{\bibfnamefont{L.~L.} \bibnamefont{Yang}}, \bibnamefont{and}
  \bibinfo{author}{\bibfnamefont{J.}~\bibnamefont{Zurita}},
  \bibinfo{journal}{JHEP} \textbf{\bibinfo{volume}{06}}, \bibinfo{pages}{016}
  (\bibinfo{year}{2013}), \eprint{1301.3492}.

\bibitem[{\citenamefont{Chetyrkin et~al.}(1998)\citenamefont{Chetyrkin, Kniehl,
  and Steinhauser}}]{Chetyrkin:1997un}
\bibinfo{author}{\bibfnamefont{K.~G.} \bibnamefont{Chetyrkin}},
  \bibinfo{author}{\bibfnamefont{B.~A.} \bibnamefont{Kniehl}},
  \bibnamefont{and}
  \bibinfo{author}{\bibfnamefont{M.}~\bibnamefont{Steinhauser}},
  \bibinfo{journal}{Nucl. Phys.} \textbf{\bibinfo{volume}{B510}},
  \bibinfo{pages}{61} (\bibinfo{year}{1998}), \eprint{hep-ph/9708255}.

\bibitem[{\citenamefont{Kramer et~al.}(1998)\citenamefont{Kramer, Laenen, and
  Spira}}]{Kramer:1996iq}
\bibinfo{author}{\bibfnamefont{M.}~\bibnamefont{Kramer}},
  \bibinfo{author}{\bibfnamefont{E.}~\bibnamefont{Laenen}}, \bibnamefont{and}
  \bibinfo{author}{\bibfnamefont{M.}~\bibnamefont{Spira}},
  \bibinfo{journal}{Nucl. Phys.} \textbf{\bibinfo{volume}{B511}},
  \bibinfo{pages}{523} (\bibinfo{year}{1998}), \eprint{hep-ph/9611272}.

\bibitem[{\citenamefont{Schroder and Steinhauser}(2006)}]{Schroder:2005hy}
\bibinfo{author}{\bibfnamefont{Y.}~\bibnamefont{Schroder}} \bibnamefont{and}
  \bibinfo{author}{\bibfnamefont{M.}~\bibnamefont{Steinhauser}},
  \bibinfo{journal}{JHEP} \textbf{\bibinfo{volume}{01}}, \bibinfo{pages}{051}
  (\bibinfo{year}{2006}), \eprint{hep-ph/0512058}.

\bibitem[{\citenamefont{Chetyrkin et~al.}(2006)\citenamefont{Chetyrkin, Kuhn,
  and Sturm}}]{Chetyrkin:2005ia}
\bibinfo{author}{\bibfnamefont{K.~G.} \bibnamefont{Chetyrkin}},
  \bibinfo{author}{\bibfnamefont{J.~H.} \bibnamefont{Kuhn}}, \bibnamefont{and}
  \bibinfo{author}{\bibfnamefont{C.}~\bibnamefont{Sturm}},
  \bibinfo{journal}{Nucl. Phys.} \textbf{\bibinfo{volume}{B744}},
  \bibinfo{pages}{121} (\bibinfo{year}{2006}), \eprint{hep-ph/0512060}.

\bibitem[{\citenamefont{Baikov and Chetyrkin}(2006)}]{Baikov:2006ch}
\bibinfo{author}{\bibfnamefont{P.~A.} \bibnamefont{Baikov}} \bibnamefont{and}
  \bibinfo{author}{\bibfnamefont{K.~G.} \bibnamefont{Chetyrkin}},
  \bibinfo{journal}{Phys. Rev. Lett.} \textbf{\bibinfo{volume}{97}},
  \bibinfo{pages}{061803} (\bibinfo{year}{2006}), \eprint{hep-ph/0604194}.

\bibitem[{\citenamefont{Inami et~al.}(1983)\citenamefont{Inami, Kubota, and
  Okada}}]{Inami:1982xt}
\bibinfo{author}{\bibfnamefont{T.}~\bibnamefont{Inami}},
  \bibinfo{author}{\bibfnamefont{T.}~\bibnamefont{Kubota}}, \bibnamefont{and}
  \bibinfo{author}{\bibfnamefont{Y.}~\bibnamefont{Okada}}, \bibinfo{journal}{Z.
  Phys.} \textbf{\bibinfo{volume}{C18}}, \bibinfo{pages}{69}
  (\bibinfo{year}{1983}).

\bibitem[{\citenamefont{Djouadi
  et~al.}(1991{\natexlab{a}})\citenamefont{Djouadi, Spira, and
  Zerwas}}]{Djouadi:1991tka}
\bibinfo{author}{\bibfnamefont{A.}~\bibnamefont{Djouadi}},
  \bibinfo{author}{\bibfnamefont{M.}~\bibnamefont{Spira}}, \bibnamefont{and}
  \bibinfo{author}{\bibfnamefont{P.~M.} \bibnamefont{Zerwas}},
  \bibinfo{journal}{Phys. Lett.} \textbf{\bibinfo{volume}{B264}},
  \bibinfo{pages}{440} (\bibinfo{year}{1991}{\natexlab{a}}).

\bibitem[{\citenamefont{Chetyrkin et~al.}(1997)\citenamefont{Chetyrkin, Kniehl,
  and Steinhauser}}]{Chetyrkin:1997iv}
\bibinfo{author}{\bibfnamefont{K.~G.} \bibnamefont{Chetyrkin}},
  \bibinfo{author}{\bibfnamefont{B.~A.} \bibnamefont{Kniehl}},
  \bibnamefont{and}
  \bibinfo{author}{\bibfnamefont{M.}~\bibnamefont{Steinhauser}},
  \bibinfo{journal}{Phys. Rev. Lett.} \textbf{\bibinfo{volume}{79}},
  \bibinfo{pages}{353} (\bibinfo{year}{1997}), \eprint{hep-ph/9705240}.

\bibitem[{\citenamefont{Anastasiou et~al.}(2015)\citenamefont{Anastasiou, Duhr,
  Dulat, Herzog, and Mistlberger}}]{Anastasiou:2015ema}
\bibinfo{author}{\bibfnamefont{C.}~\bibnamefont{Anastasiou}},
  \bibinfo{author}{\bibfnamefont{C.}~\bibnamefont{Duhr}},
  \bibinfo{author}{\bibfnamefont{F.}~\bibnamefont{Dulat}},
  \bibinfo{author}{\bibfnamefont{F.}~\bibnamefont{Herzog}}, \bibnamefont{and}
  \bibinfo{author}{\bibfnamefont{B.}~\bibnamefont{Mistlberger}},
  \bibinfo{journal}{Phys. Rev. Lett.} \textbf{\bibinfo{volume}{114}},
  \bibinfo{pages}{212001} (\bibinfo{year}{2015}), \eprint{1503.06056}.

\bibitem[{\citenamefont{Spira et~al.}(1995)\citenamefont{Spira, Djouadi,
  Graudenz, and Zerwas}}]{Spira:1995rr}
\bibinfo{author}{\bibfnamefont{M.}~\bibnamefont{Spira}},
  \bibinfo{author}{\bibfnamefont{A.}~\bibnamefont{Djouadi}},
  \bibinfo{author}{\bibfnamefont{D.}~\bibnamefont{Graudenz}}, \bibnamefont{and}
  \bibinfo{author}{\bibfnamefont{P.~M.} \bibnamefont{Zerwas}},
  \bibinfo{journal}{Nucl. Phys.} \textbf{\bibinfo{volume}{B453}},
  \bibinfo{pages}{17} (\bibinfo{year}{1995}), \eprint{hep-ph/9504378}.

\bibitem[{\citenamefont{Djouadi et~al.}(1993)\citenamefont{Djouadi, Spira, and
  Zerwas}}]{Djouadi:1993ji}
\bibinfo{author}{\bibfnamefont{A.}~\bibnamefont{Djouadi}},
  \bibinfo{author}{\bibfnamefont{M.}~\bibnamefont{Spira}}, \bibnamefont{and}
  \bibinfo{author}{\bibfnamefont{P.~M.} \bibnamefont{Zerwas}},
  \bibinfo{journal}{Phys. Lett.} \textbf{\bibinfo{volume}{B311}},
  \bibinfo{pages}{255} (\bibinfo{year}{1993}), \eprint{hep-ph/9305335}.

\bibitem[{\citenamefont{Zheng and Wu}(1990)}]{Zheng:1990qa}
\bibinfo{author}{\bibfnamefont{H.-Q.} \bibnamefont{Zheng}} \bibnamefont{and}
  \bibinfo{author}{\bibfnamefont{D.-D.} \bibnamefont{Wu}},
  \bibinfo{journal}{Phys. Rev.} \textbf{\bibinfo{volume}{D42}},
  \bibinfo{pages}{3760} (\bibinfo{year}{1990}).

\bibitem[{\citenamefont{Djouadi
  et~al.}(1991{\natexlab{b}})\citenamefont{Djouadi, Spira, van~der Bij, and
  Zerwas}}]{Djouadi:1990aj}
\bibinfo{author}{\bibfnamefont{A.}~\bibnamefont{Djouadi}},
  \bibinfo{author}{\bibfnamefont{M.}~\bibnamefont{Spira}},
  \bibinfo{author}{\bibfnamefont{J.~J.} \bibnamefont{van~der Bij}},
  \bibnamefont{and} \bibinfo{author}{\bibfnamefont{P.~M.}
  \bibnamefont{Zerwas}}, \bibinfo{journal}{Phys. Lett.}
  \textbf{\bibinfo{volume}{B257}}, \bibinfo{pages}{187}
  (\bibinfo{year}{1991}{\natexlab{b}}).

\bibitem[{\citenamefont{Dawson and Kauffman}(1993)}]{Dawson:1992cy}
\bibinfo{author}{\bibfnamefont{S.}~\bibnamefont{Dawson}} \bibnamefont{and}
  \bibinfo{author}{\bibfnamefont{R.~P.} \bibnamefont{Kauffman}},
  \bibinfo{journal}{Phys. Rev.} \textbf{\bibinfo{volume}{D47}},
  \bibinfo{pages}{1264} (\bibinfo{year}{1993}).

\bibitem[{\citenamefont{Melnikov and Yakovlev}(1993)}]{Melnikov:1993tj}
\bibinfo{author}{\bibfnamefont{K.}~\bibnamefont{Melnikov}} \bibnamefont{and}
  \bibinfo{author}{\bibfnamefont{O.~I.} \bibnamefont{Yakovlev}},
  \bibinfo{journal}{Phys. Lett.} \textbf{\bibinfo{volume}{B312}},
  \bibinfo{pages}{179} (\bibinfo{year}{1993}), \eprint{hep-ph/9302281}.

\bibitem[{\citenamefont{Inoue et~al.}(1994)\citenamefont{Inoue, Najima, Oka,
  and Saito}}]{Inoue:1994jq}
\bibinfo{author}{\bibfnamefont{M.}~\bibnamefont{Inoue}},
  \bibinfo{author}{\bibfnamefont{R.}~\bibnamefont{Najima}},
  \bibinfo{author}{\bibfnamefont{T.}~\bibnamefont{Oka}}, \bibnamefont{and}
  \bibinfo{author}{\bibfnamefont{J.}~\bibnamefont{Saito}},
  \bibinfo{journal}{Mod. Phys. Lett.} \textbf{\bibinfo{volume}{A9}},
  \bibinfo{pages}{1189} (\bibinfo{year}{1994}).

\bibitem[{\citenamefont{Fleischer et~al.}(2004)\citenamefont{Fleischer,
  Tarasov, and Tarasov}}]{Fleischer:2004vb}
\bibinfo{author}{\bibfnamefont{J.}~\bibnamefont{Fleischer}},
  \bibinfo{author}{\bibfnamefont{O.~V.} \bibnamefont{Tarasov}},
  \bibnamefont{and} \bibinfo{author}{\bibfnamefont{V.~O.}
  \bibnamefont{Tarasov}}, \bibinfo{journal}{Phys. Lett.}
  \textbf{\bibinfo{volume}{B584}}, \bibinfo{pages}{294} (\bibinfo{year}{2004}),
  \eprint{hep-ph/0401090}.

\bibitem[{\citenamefont{Aglietti
  et~al.}(2004{\natexlab{a}})\citenamefont{Aglietti, Bonciani, Degrassi, and
  Vicini}}]{Aglietti:2004nj}
\bibinfo{author}{\bibfnamefont{U.}~\bibnamefont{Aglietti}},
  \bibinfo{author}{\bibfnamefont{R.}~\bibnamefont{Bonciani}},
  \bibinfo{author}{\bibfnamefont{G.}~\bibnamefont{Degrassi}}, \bibnamefont{and}
  \bibinfo{author}{\bibfnamefont{A.}~\bibnamefont{Vicini}},
  \bibinfo{journal}{Phys. Lett.} \textbf{\bibinfo{volume}{B595}},
  \bibinfo{pages}{432} (\bibinfo{year}{2004}{\natexlab{a}}),
  \eprint{hep-ph/0404071}.

\bibitem[{\citenamefont{Aglietti
  et~al.}(2004{\natexlab{b}})\citenamefont{Aglietti, Bonciani, Degrassi, and
  Vicini}}]{Aglietti:2004ki}
\bibinfo{author}{\bibfnamefont{U.}~\bibnamefont{Aglietti}},
  \bibinfo{author}{\bibfnamefont{R.}~\bibnamefont{Bonciani}},
  \bibinfo{author}{\bibfnamefont{G.}~\bibnamefont{Degrassi}}, \bibnamefont{and}
  \bibinfo{author}{\bibfnamefont{A.}~\bibnamefont{Vicini}},
  \bibinfo{journal}{Phys. Lett.} \textbf{\bibinfo{volume}{B600}},
  \bibinfo{pages}{57} (\bibinfo{year}{2004}{\natexlab{b}}),
  \eprint{hep-ph/0407162}.

\bibitem[{\citenamefont{Actis et~al.}(2009)\citenamefont{Actis, Passarino,
  Sturm, and Uccirati}}]{Actis:2008ts}
\bibinfo{author}{\bibfnamefont{S.}~\bibnamefont{Actis}},
  \bibinfo{author}{\bibfnamefont{G.}~\bibnamefont{Passarino}},
  \bibinfo{author}{\bibfnamefont{C.}~\bibnamefont{Sturm}}, \bibnamefont{and}
  \bibinfo{author}{\bibfnamefont{S.}~\bibnamefont{Uccirati}},
  \bibinfo{journal}{Nucl. Phys.} \textbf{\bibinfo{volume}{B811}},
  \bibinfo{pages}{182} (\bibinfo{year}{2009}), \eprint{0809.3667}.

\bibitem[{\citenamefont{Degrassi and Maltoni}(2004)}]{Degrassi:2004mx}
\bibinfo{author}{\bibfnamefont{G.}~\bibnamefont{Degrassi}} \bibnamefont{and}
  \bibinfo{author}{\bibfnamefont{F.}~\bibnamefont{Maltoni}},
  \bibinfo{journal}{Phys. Lett.} \textbf{\bibinfo{volume}{B600}},
  \bibinfo{pages}{255} (\bibinfo{year}{2004}), \eprint{hep-ph/0407249}.

\bibitem[{\citenamefont{Actis et~al.}(2008)\citenamefont{Actis, Passarino,
  Sturm, and Uccirati}}]{Actis:2008ug}
\bibinfo{author}{\bibfnamefont{S.}~\bibnamefont{Actis}},
  \bibinfo{author}{\bibfnamefont{G.}~\bibnamefont{Passarino}},
  \bibinfo{author}{\bibfnamefont{C.}~\bibnamefont{Sturm}}, \bibnamefont{and}
  \bibinfo{author}{\bibfnamefont{S.}~\bibnamefont{Uccirati}},
  \bibinfo{journal}{Phys. Lett.} \textbf{\bibinfo{volume}{B670}},
  \bibinfo{pages}{12} (\bibinfo{year}{2008}), \eprint{0809.1301}.

\bibitem[{\citenamefont{Degrassi and Maltoni}(2005)}]{Degrassi:2005mc}
\bibinfo{author}{\bibfnamefont{G.}~\bibnamefont{Degrassi}} \bibnamefont{and}
  \bibinfo{author}{\bibfnamefont{F.}~\bibnamefont{Maltoni}},
  \bibinfo{journal}{Nucl. Phys.} \textbf{\bibinfo{volume}{B724}},
  \bibinfo{pages}{183} (\bibinfo{year}{2005}), \eprint{hep-ph/0504137}.

\bibitem[{\citenamefont{Bernon and Dumont}(2015)}]{Bernon:2015hsa}
\bibinfo{author}{\bibfnamefont{J.}~\bibnamefont{Bernon}} \bibnamefont{and}
  \bibinfo{author}{\bibfnamefont{B.}~\bibnamefont{Dumont}},
  \bibinfo{journal}{Eur. Phys. J.} \textbf{\bibinfo{volume}{C75}},
  \bibinfo{pages}{440} (\bibinfo{year}{2015}), \eprint{1502.04138}.

\bibitem[{\citenamefont{ATLAS and Collaborations}(2015)}]{ATLAS:2015}
\bibinfo{author}{\bibfnamefont{T.}~\bibnamefont{ATLAS}} \bibnamefont{and}
  \bibinfo{author}{\bibfnamefont{C.}~\bibnamefont{Collaborations}},
  \bibinfo{journal}{ATLAS-CONF-2015-044}  (\bibinfo{year}{2015}).

\bibitem[{\citenamefont{Collaboration}(2015)}]{CMS:2015kwa}
\bibinfo{author}{\bibfnamefont{C.}~\bibnamefont{Collaboration}}
  (\bibinfo{collaboration}{CMS}), \bibinfo{journal}{CMS-PAS-HIG-15-002}
  (\bibinfo{year}{2015}).

\bibitem[{\citenamefont{Aaltonen et~al.}(2013)}]{Aaltonen:2013ioz}
\bibinfo{author}{\bibfnamefont{T.}~\bibnamefont{Aaltonen}} \bibnamefont{et~al.}
  (\bibinfo{collaboration}{CDF, D0}), \bibinfo{journal}{Phys. Rev.}
  \textbf{\bibinfo{volume}{D88}}, \bibinfo{pages}{052014}
  (\bibinfo{year}{2013}), \eprint{1303.6346}.

\bibitem[{\citenamefont{Gemme}(2016)}]{Gemme:2016wgs}
\bibinfo{author}{\bibfnamefont{C.}~\bibnamefont{Gemme}} (\bibinfo{year}{2016}),
  \eprint{1612.01987}.

\bibitem[{\citenamefont{Palencia~Cortezon}(2016)}]{PalenciaCortezon:2016yiz}
\bibinfo{author}{\bibfnamefont{E.}~\bibnamefont{Palencia~Cortezon}}
  (\bibinfo{collaboration}{CMS}), in \emph{\bibinfo{booktitle}{{9th
  International Workshop on Top Quark Physics (TOP 2016) Olomouc, Czech
  Republic, September 19-23, 2016}}} (\bibinfo{year}{2016}),
  \eprint{1612.00946},
  \urlprefix\url{http://inspirehep.net/record/1501679/files/arXiv:1612.00946.pdf}.

\bibitem[{\citenamefont{Buchalla et~al.}(2016)\citenamefont{Buchalla, Cata,
  Celis, and Krause}}]{Buchalla:2015qju}
\bibinfo{author}{\bibfnamefont{G.}~\bibnamefont{Buchalla}},
  \bibinfo{author}{\bibfnamefont{O.}~\bibnamefont{Cata}},
  \bibinfo{author}{\bibfnamefont{A.}~\bibnamefont{Celis}}, \bibnamefont{and}
  \bibinfo{author}{\bibfnamefont{C.}~\bibnamefont{Krause}},
  \bibinfo{journal}{Eur. Phys. J.} \textbf{\bibinfo{volume}{C76}},
  \bibinfo{pages}{233} (\bibinfo{year}{2016}), \eprint{1511.00988}.

\end{thebibliography}

\end{document}